\documentclass[aip,jap,reprint]{revtex4-1}

\usepackage{amsmath}
\usepackage{graphicx}
\usepackage{units}
\usepackage{color}

\graphicspath{{./fig/}}

\begin{document}

\title{Analysis of slope-intercept plots for arrays of electron field emitters}
\author{Arun Persaud}
\email{APersaud@lbl.gov}
\affiliation{E.O. Lawrence Berkeley National Laboratory, Berkeley, CA 94720, USA}

\begin{abstract}
  In electron field emission experiments, a linear relationship in
  plots of slope vs. intercept obtained from Fowler-Nordheim analysis is
  commonly observed for single tips or tip arrays. By simulating
  samples with many tips, it is shown here that the observed linear
  relationship results from the distribution of input parameters,
  assuming a log-normal distribution for the radius of
  each tip. Typically, a shift from the lower-left to the upper-right
  of a slope-intercept plot has been correlated with a shift in
  work function. However, as shown in this paper, the same effect can
  result from a variation in the number of emitters.
\end{abstract}

\maketitle

\section{Introduction}

Field emitters such as carbon nanofibers play an important role in
today's vacuum electronics.\cite{Gaertner2012} They are also used
in ion sources, for example field ionization sources for neutron generators.
\cite{Naranjo2005,Resnick20101263, Ellsworth2013, Persaud:RoSI-83-02B312,
  Persaud:JVSTB-29-02B107} In all cases, it is important to be able to
characterize the field emitter samples, which can be done using
electron field emission experiments, see for example Ref.~\onlinecite{Waldmann:TSF-534-48}.

Electron field emission occurs as a result of a tunneling process of
electrons through a potential barrier. This was first reported by
Fowler and Nordheim in 1928,\cite{Fowler1928} who derived an equation
for the emitted current. In their 1-dimensional model, the emission
current is expressed as a function of the local field, the emitter
area, and the work function of the material. Since then, the model has
been extended to include a number of correction factors.\cite{Forbes2013}

The Fowler-Nordheim (FN) equation can be rewritten using a different set of
variables, such that the relationship between these new variables is
linear. The linear fit results in two parameters: the slope and the intercept. If
one acquires several FN-plots for the same sample or for a set of
different samples, different slope-intercept data points are
obtained. When plotting the slope against the intercept, an almost
linear relationship has been observed in many cases.
\cite{Mackie2003, Charbonnier2004, Gotoh2007a, Kawasaki2010} It has not been clear
what the origin of this linear relationship is or what it
implies. In this paper, the linearity is explained using a
statistical model simulating field emission from tips with a random
distribution of height, radius, and work function.

\section{Theory}

In the following, the equations used for the simulations presented in this paper
are introduced.  For an introduction into the theory of
field emission, the reader is referred to Gomer, \cite{Gomer1994} and for a
short summary to the article on FN-analysis by Forbes \textit{et al.}\cite{Forbes2012,Forbes2013}

The simplest equation for FN-tunneling is given by
\begin{equation}
  \label{eq:FN-simple}
I = \frac{c_a A E_{local}^2}{\Phi} \exp\left(\frac{-c_b \Phi^{1.5}}{E_{local}}\right),
\end{equation}

where $I$ is the emitted current, $c_a$ and $c_b$ are the so-called
first and second FN-constants, $A$ is the emitting area, $\Phi$ is the
work function of the material, and $E_{local}$ is the local electric
field. The values for the FN-constants are
$c_a=\frac{e^3}{16\pi^2\hbar}=\unitfrac[1.5414\times10^{-6}]{A
  eV}{V^2}$ and
$c_b=\frac{4\sqrt{2m_e}}{3e\hbar}=\unitfrac[6.8309\times10^9]{V
  m}{eV^{1.5}}$, with $e$, $m_e$, being the charge and mass of an
electron, respectively, and $\hbar$ being the reduced Planck constant.

The local field can be expressed using the applied field $E$ and a local
field enhancement factor $\gamma$, as
\begin{equation}
  \label{eq:gamma}
  E_{local}=\gamma E.
\end{equation}

Equation~\eqref{eq:FN-simple} is derived for the 1-dimensional case and a
uniform applied field.  In experiments, the field is enhanced at
a step or a sharp point.  The enhancement factor $\gamma$ directly expresses
the amount of enhancement from a tip-like structure compared
to a flat surface.  For cylindrical tips, a simple estimate
is given by
\begin{equation}
  \label{eq:gamma-simple}
  \gamma_{simple} = \frac{H}{R},
\end{equation}
where $H$ is the height of the tip and $R$ the radius. A more
precise estimate is given by Edgcombe and Valdr\`e \cite{Edgcombe2001}
as
\begin{equation}
  \label{eq:gamma-complex}
  \gamma = 1.2\left[\frac{H}{R}+2.5\right]^{0.9}.
\end{equation}
If there are many emitters in an array structure, there is also a
shielding effect that needs to be taken into account, especially if
the tips are very close to each other. The shielding effect can be
estimated \cite{Bonard2001,Jo2003} using
\begin{equation}
  \label{eq:gamma-array}
  \gamma_{array}=\gamma \left[1-\exp\left(- 2.3171\frac{S}{H}\right)\right],
\end{equation}
where $S$ is the spacing between the tips.

It can be helpful to rescale the FN-equation using
\begin{equation}
  \label{eq:rescale}
  f = \frac{c_S^2}{\Phi^2} E_{local},
\end{equation}
where $c_S$ is the Schottky constant ($c_S^2=\frac{e^3}{4 \pi
  \epsilon_0}$; $\epsilon_0$ is the electric constant). By doing
this, the FN-equation now depends on a dimensionless variable. Using this
rescaling approach and adding some correction factors, the FN-equation
can be written as \cite{Forbes2013}
\begin{equation}
  \label{eq:FN-complex}
  I = A c_a \frac{\Phi^3}{c_S^4}f^2 \exp\left(-\frac{c_b c_S^2}{\sqrt{\Phi}}\left[1-f+\frac{1}{6}f\ln(f)\right]\frac{1}{f}\right).
\end{equation}
This equation is physically more realistic than
Eq.~\eqref{eq:FN-simple}, since it takes image forces into
account. This has the effect that the tunneling barrier is lower
compared to the case in Eq.~\eqref{eq:FN-simple},
and higher emission currents (typically by a factor of 100 or more)
are predicted.

The linear relationship of the FN-plot is seen when rewriting Eq.~\eqref{eq:FN-simple} as
\begin{equation}
  \label{eq:FN}
  \ln\left(\frac{I}{E^2}\right) = \ln\left(\frac{c_a A \gamma^2}{\Phi}\right) -
  \frac{c_b \Phi^{1.5}}{\gamma} \frac{1}{E}.
\end{equation}
Clearly, this equation has the linear form,
\begin{equation}
  \label{eq:linear}
  \ln\left(\frac{I}{E^2}\right) = b + m \frac{1}{E}.
\end{equation}
More realistic equations, such as equation~\eqref{eq:FN-complex}, generate nearly (but not exactly) linear FN-plots.
These more realistic theoretical plots, and also many experimental FN-plots, can in practice be
fitted by a linear equation of form Eq.~\eqref{eq:linear}. In all the figures in this paper, the quantity $\frac{I}{E^2}$ is
expressed in the units ($\unit{A V^{-2} m^2}$) before the natural logarithm is taken.

From the slope $m$ and the
intercept $b$ of the fit, one can calculate two of the three parameters
$A$, $\Phi$, and $\gamma$, if one makes an assumption about the third
parameter. Often one assumes a known value for the work function in
order to
extract the other two parameters. Due to the form of the equation,
it is not easy to obtain information on the work function and the
field enhancement factor at the same time, since they appear almost as
$\frac{\Phi}{\gamma}$, which makes them difficult to distinguish. Furthermore, it is non-trivial to define the emitting
area. This is due to the fact that the original equation was derived
for the 1-dimensional case, whereas for a 3-dimensional sharp tip the
electric field will not be constant across the surface
and there will be a field dependence of the area that is not included
in these equations. In addition, time dependence of any of the parameters is
not included in this analysis. Time dependent effects include
changes in the work function and field enhancement factors due to
absorbents on the surface or blunting of the tips due to sputtering or
heating effects.

\section{Slope-intercept plots}
\label{seq:SK}

In order to compare different samples or to examine the behavior of a single sample
over time, plotting the slope and intercepts from fitting different
FN-plots was introduced.\cite{Ishikawa1993,Gotoh1996} Often these
plots are also referred to using their Japanese names: Seppen-Katamuki plots
or SK-plots for short. From Eq.~\eqref{eq:FN-simple}, one can calculate lines of constant
work function or constant field enhancement factor for these
plots. Assuming one of the relationships between radius, height, and field
enhancement factors listed above, equations for constant height and constant radius
can also be derived.\cite{Ishikawa1993,Gotoh1996,Mackie2003,DaRocha2008,Kawasaki2010,Nicolaescu2001}

In SK-plots, a linear relationship between the slope and the
intercept is often observed.
\cite{Gotoh2007b,Ishikawa1993,Gotoh1996,Kawasaki2010} However, this
linear relationship cannot easily be explained assuming only a
change in a single parameter, such as the radius. For a possible explanation using the simple FN-equation,
one needs to assume a more complex radial dependency of the
area.\cite{Ishikawa1993,Gotoh2001}  Another problem with explanations
of this kind can be seen if one looks at data from repeated
experiments, for example, as shown in Figure~7 in the paper of Gotoh
\textit{et al.}\cite{Gotoh2004} The data points do not start at one end of the line
and then move along to the other end, as one would expect if the change
would be due to a continuous function, but rather the
distribution of data points along the line appears to be random.

The linear relation shown in the SK-plots seems to be orthogonal to
a change in work function and therefore different parallel lines in a
SK-plot have been attributed to a change in the work function and absolute
values have been extracted. A good example of this can
be seen in the work of Gotoh \textit{et al.},\cite{Gotoh2007b} where
the authors were able to extract the work functions for different
crystal planes of a tungsten emitter.

Charbonnier, Southall, and Mackie also investigated further possible
explanations for the linear behavior in SK-plots, and considered whether
nano-protrusions on top of a single emitter can be the cause. They found
that a single protrusion cannot fit the experimental data, but if
several protrusions are used then good linear fits to the SK-plots can
be obtained by choosing the correct parameters for these protrusions.
\cite{Charbonnier2004}

\section{Simplified model}

For an array of $N$ tips, the FN-equation becomes more complicated,
since one has to deal with many tips that will show a distribution of
heights, radii and work functions. Also, the shielding
between tips cannot be easily computed, since neighboring tips
will have different heights and radii and therefore the shielding
effect will vary. The shielding effect can be neglected, however,
if the distance between tips is larger than twice the height of the
tips.\cite{Bonard2001} From here on, the assumption is
made that the tips on the sample are
spaced even further apart and therefore any shielding
effect is ignored. To make the model very simple, Eqs.~\eqref{eq:FN-simple}--
\eqref{eq:gamma-simple} are used. The index $i$ is used to distinguish
different tips, that is, $R_i$ will be the radius of tip $i$,
etc. Furthermore, a very basic relation for the area of
each tip, $A_i=\pi r_i^2$, is assumed. The total current for a single sample can
then be written as
\begin{equation}
  \label{eq:sample}
  I = \sum_i^N \frac{c_a \pi H_i^2 E^2}{\Phi_i} \exp\left(\frac{-c_b \Phi_i^{1.5}R_i}{H_iE}\right).
\end{equation}

Nicolaescu \textit{et al.} have used a similar equation to directly fit
FN-plots using parameters of the distribution as fitting parameters.
\cite{Nicolaescu2003,Nicolaescu2004}

The distribution of most parameters is most likely a log-normal
distribution (negative values for height and radius from a
normal Gaussian distribution would not make sense). A
log-normal distribution for the radius has also been observed by Ding
\textit{et al.},\cite{Ding2002,Nicolaescu2006} as well as by Park \textit{et al.} for
the field enhancement factor.\cite{Park2006}

\section{Simulation results}

\begin{figure}[ht!]
  \centering
  \includegraphics[width=\linewidth]{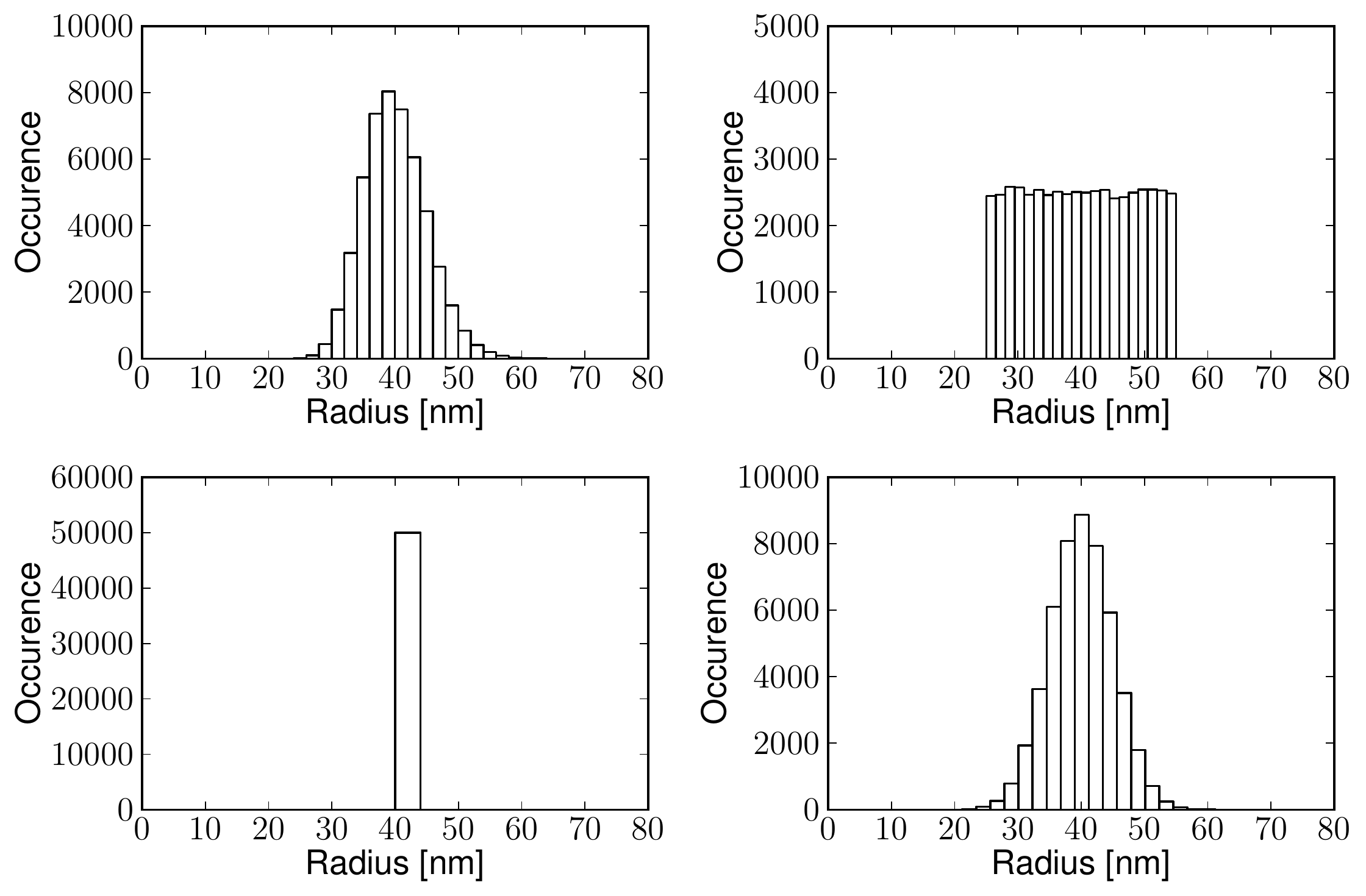}
  \caption{The investigated distributions using the radial
    distribution as an example. Top row: log-normal distribution, uniform distribution,
     bottom row: constant value, Gaussian distribution.}
  \label{fig:distributions}
\end{figure}

In the following, the results of simulations using Eq.~\eqref{eq:sample}
with varying parameters are shown. The difference between a simple model and
using the more realistic FN-equation~\eqref{eq:FN-complex}, and the
different approximations for the field enhancement factors are also
examined. Various random distributions are explored (log-normal,
normal, uniform), as well as the effect of just using a constant value. When comparing a
log-normal/Gaussian distribution with a uniform distribution, the range
in values for the uniform distribution is chosen to coincide with six standard
deviations of the log-normal/Gaussian distribution. Histograms of
example distributions are plotted in Fig.~\ref{fig:distributions}.

The simulations were carried out using
\textsc{python}.\cite{python} The numerical python module numpy\cite{Jones2001} was used
for most of the calculations and for random number generation,
pylab\cite{Hunter:2007} was used for plotting, scipy\cite{Jones2001}
was used for the fitting routine, and ipython\cite{PER-GRA:2007} was
utilized to execute the main routine in parallel on a local
ipcluster.  The basic algorithm consisted of the following steps for
calculating $K$ samples of size $S$: a) generate $S$ random
distributed values for radius, height and work function, and calculate
field enhancement factors for each tip; b) for a given range of applied electric
field, calculate the total current for the sample for each field value;
c) perform a linear fit on the simulated data using Eq.~\eqref{eq:linear} and save
the fit parameters; d) repeat steps a)-c) $K$ times to generate
statistics; e) plot results.

For the following results, the values shown in Table~\ref{tab:values}
are used unless otherwise noted. The electric fields used for the fit were
20 evenly spaced values between $\unitfrac[1.6]{V}{\mu
  m}$ and $\unitfrac[26]{V}{\mu m}$.
\begin{table}[ht!]
  \centering
  \begin{tabular}{cccc}
    Parameter          & Radius        & Height              & Work function \\ \hline \hline
    Average $\mu$      & \unit[40]{nm} & \unit[6]{$\mu$m}    & \unit[4.8]{eV} \\
    Deviation $\sigma$ & \unit[5]{nm}  & \unit[0.1]{$\mu$m}  & \unit[0.3]{eV}\\ \hline
  \end{tabular}
  \caption{Parameters used in the simulation. For a uniform distribution,
     the minimum and maximum values are taken as $\mu \pm
    3\sigma$ and for constant values just the value for $\mu$ is used.}
  \label{tab:values}
\end{table}

\begin{figure}[ht!]
  \centering
  \includegraphics[width=\linewidth]{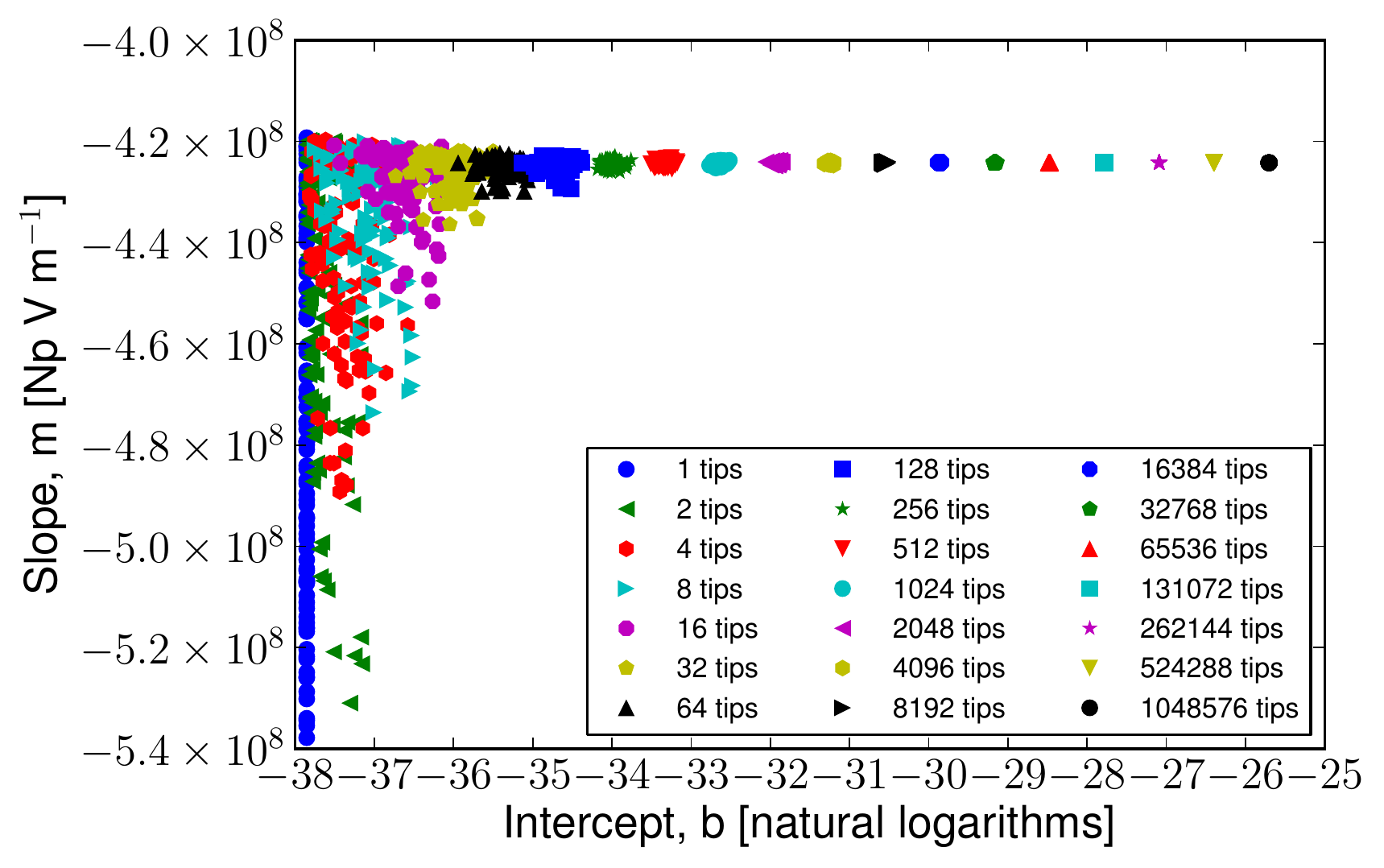}
  \caption{SK-plot for uniform radial distribution, constant height and work
    function, variation in sample size. For each sample size 100
    samples have been simulated.}
  \label{fig:uniform_R}
\end{figure}

\begin{figure}[ht!]
  \centering
  \includegraphics[width=\linewidth]{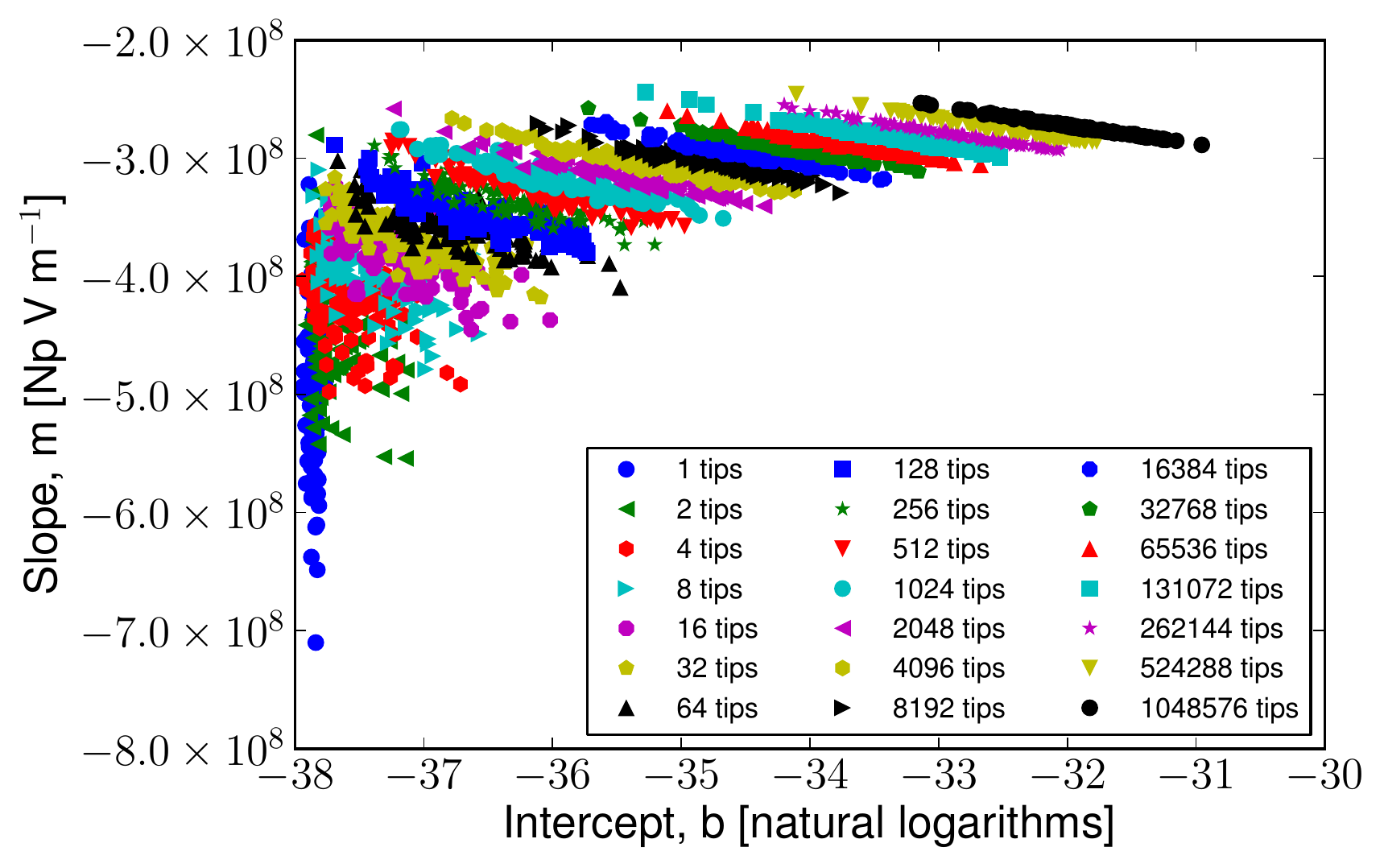}
  \caption{SK-plot for log-normal radial distribution, constant height and work
    function, variation in sample size. For each sample size, 100
    samples have been simulated.}
  \label{fig:sizes}
\end{figure}

\begin{figure}[ht!]
  \centering
  \includegraphics[width=\linewidth]{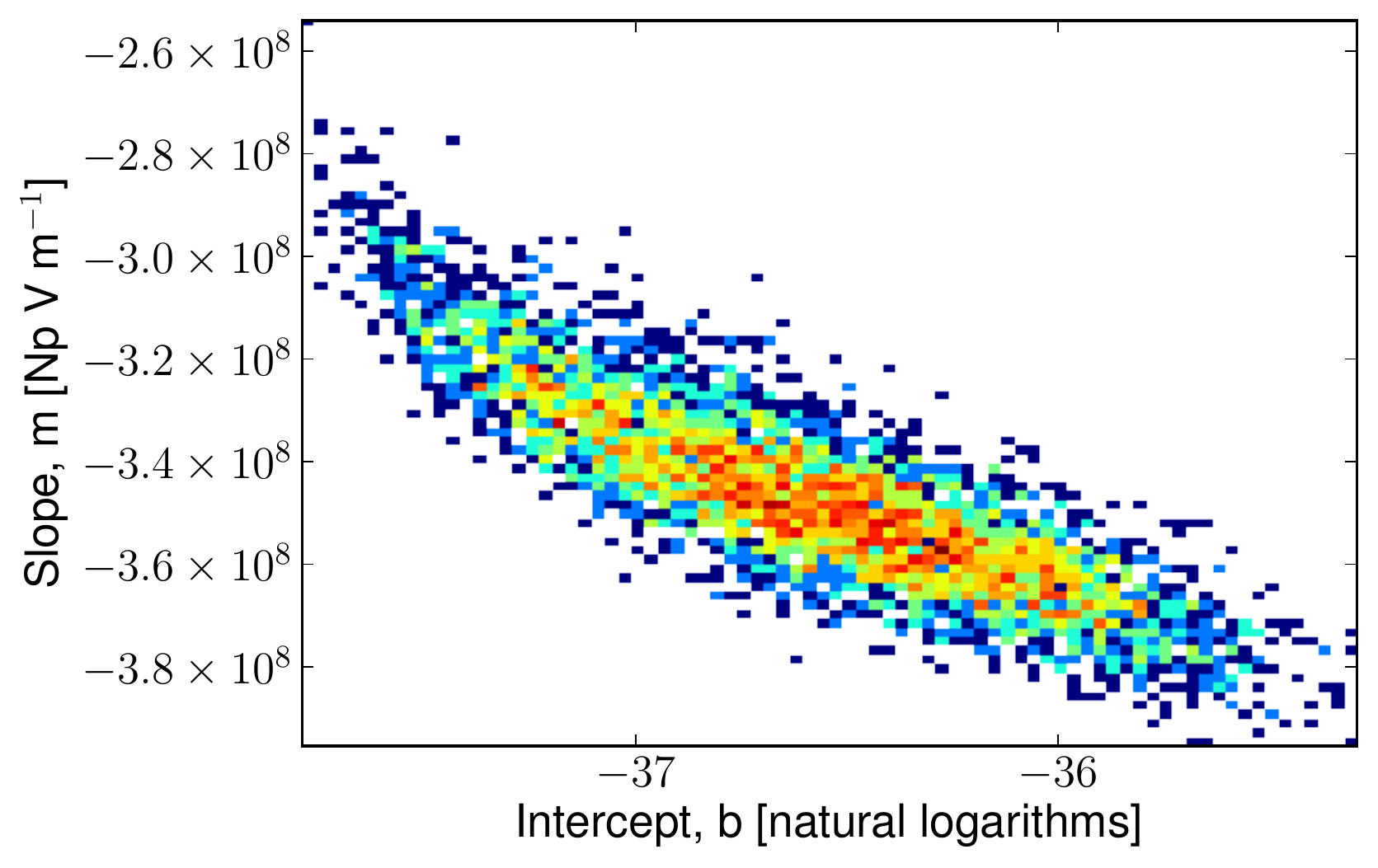}
  \caption{SK-plot for log-normal radial distribution, log-normal height, and constant work
    function, sample size 128. 5000 samples have been simulated.}
  \label{fig:zoom}
\end{figure}

In Figs.~\ref{fig:uniform_R} and \ref{fig:sizes}, results are plotted
for a sample with a constant height and a constant work function, but
different radial distributions: uniform and
log-normal.  Furthermore, the number of tips per sample, $N$ in
Eq.~\eqref{eq:sample}, is varied as shown in the plots.  One can see
that a nearly linear relationship emerges for large $N$ and a
log-normal distribution of tip radii.
This emerging behavior is purely statistical in nature due to the fact
that a log-normal distribution has a non-zero probability of
containing very sharp tips. However, when one looks more closely at a
single SK-plot and generates better statistics, a different picture
is revealed. A histogram of a simulation using 128 emitters per sample
for a total of 5000 samples is shown in Fig.~\ref{fig:zoom}
and shows that instead of a linear relationship the SK-plot is
actually closer to a 2-dimensional Gaussian distribution, which would
be expected for large numbers. From Fig.~\ref{fig:sizes}, it is clear
that the greater the number of tips/sample the narrower the
2-dimensional Gaussian distribution will be, resulting in the almost
linear appearance.

\begin{figure}[ht!]
  \centering
  \includegraphics[width=\linewidth]{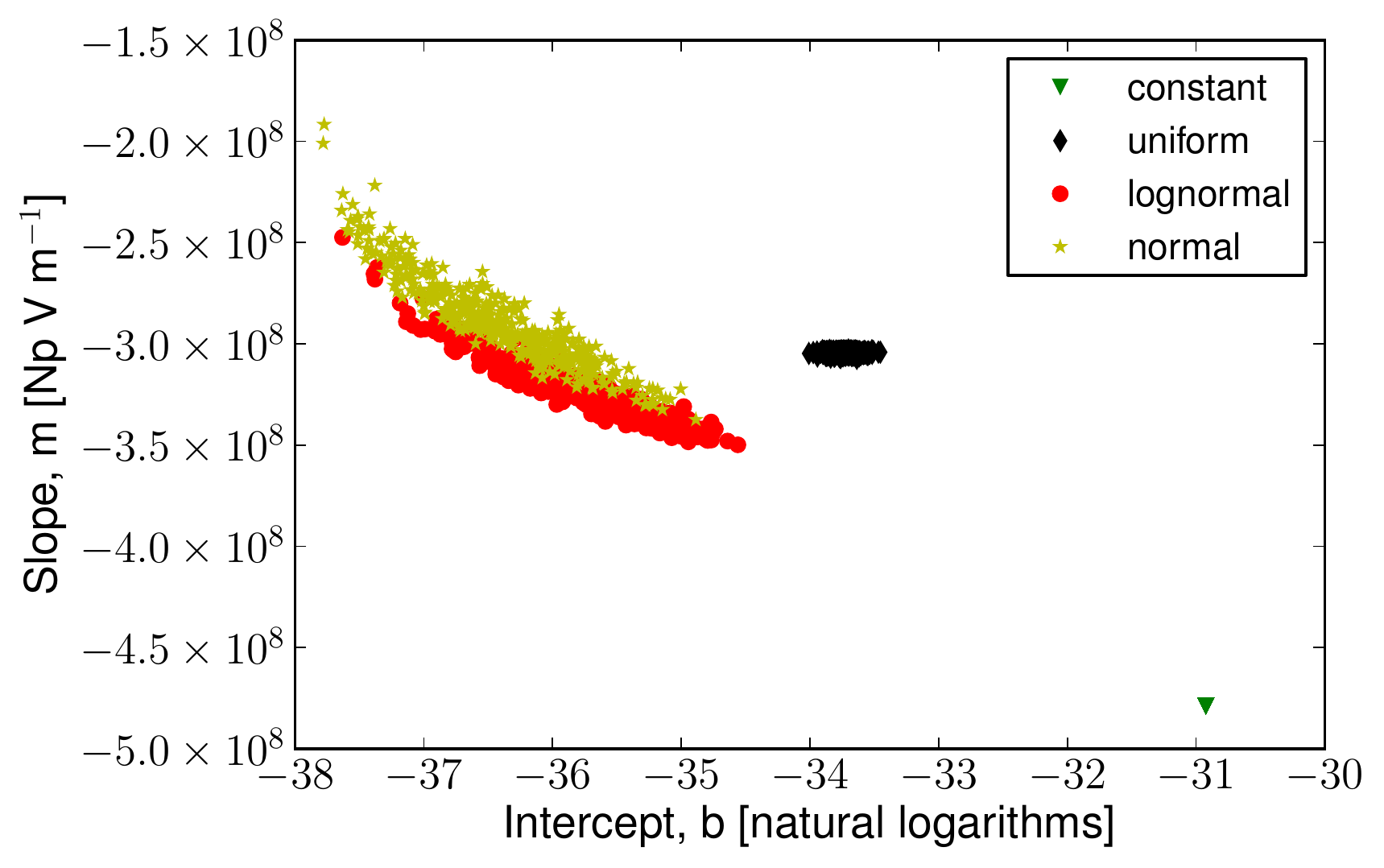}
  \caption{SK-plot for constant height and work function with varying
    distributions for the radius. Sample size 1024 tips, 400 samples
    have been simulated for each distribution.}
  \label{fig:Rdist}
\end{figure}

In Fig~\ref{fig:Rdist}, the dependency of the
shape of the SK-plot on the type of radial distribution is
explored. Obviously, if only constant values are used, all data points in the SK-plot are the same. A
uniform distribution for the radius results in a horizontal
distribution of points that only varies slightly on the SK-plot. This
is similar to plots of constant radius in previous publications,
although using a simpler model for the dependency of the FN-equation on
tip radius. A log-normal or Gaussian distribution, however, shows an
almost linear spread across a wide range of values in the SK-plot,
very similar to experimental observations.

\begin{figure}[ht!]
  \centering
  \includegraphics[width=\linewidth]{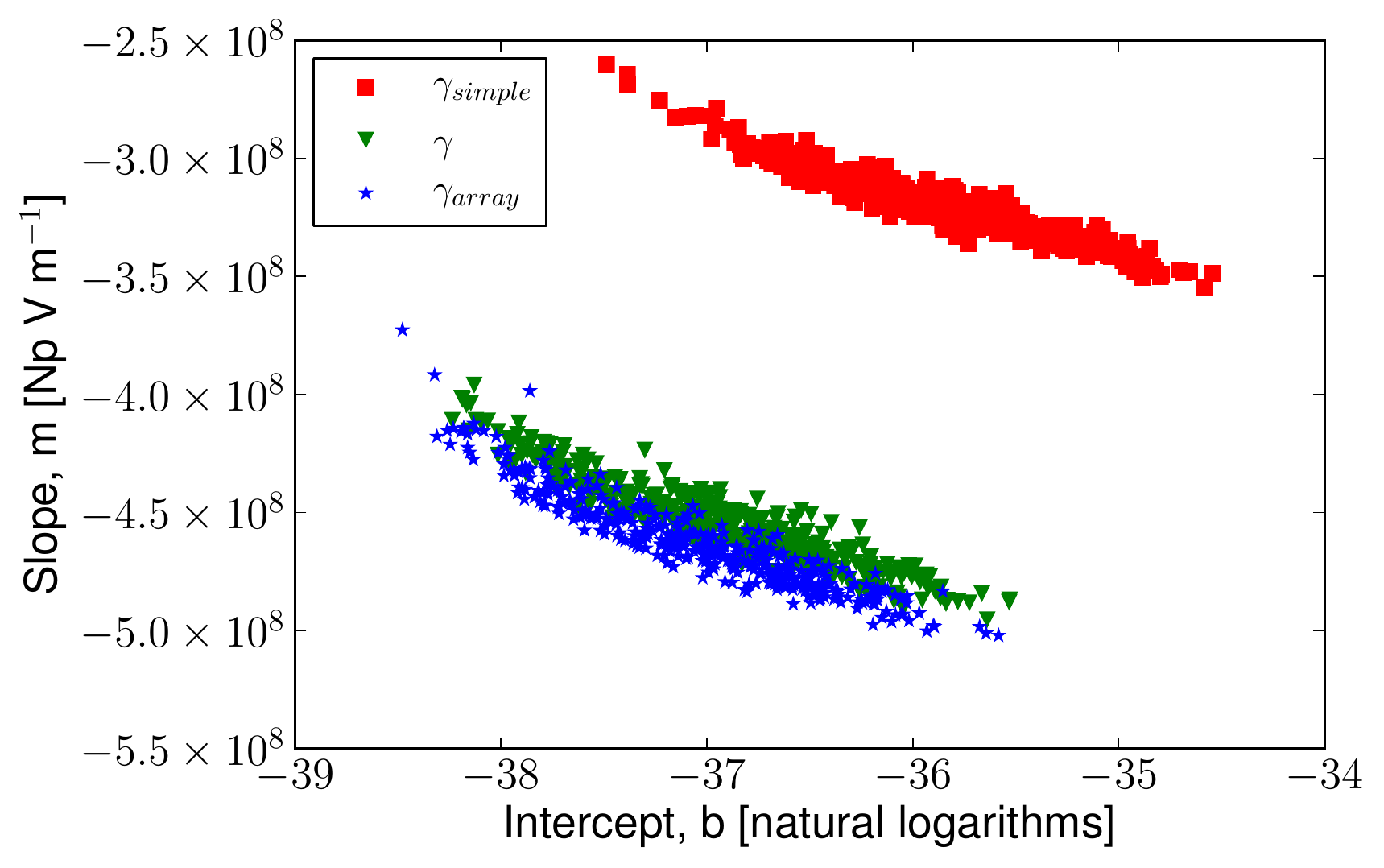}
  \caption{SK-plot for log-normal radial distribution, log-normal
    height, and constant work function. Comparing the effect of
    different approximations on the field enhancement factor. For the
    array simulation a spacing of $\unit[10]{\mu m}$ was assumed.}
  \label{fig:gamma}
\end{figure}

It is also easy to simulate the effect of the different approximations
for the field enhancement factor
(Eqs.~\eqref{eq:gamma-simple}--\eqref{eq:gamma-array}). The results
are shown in Fig.~\ref{fig:gamma}. Although there is a shift in
values, the shape of the resulting plot does not change.  Therefore,
for the purpose of this paper the use of the simplest approximation is
justified.

\begin{figure}[ht!]
  \centering
  \includegraphics[width=\linewidth]{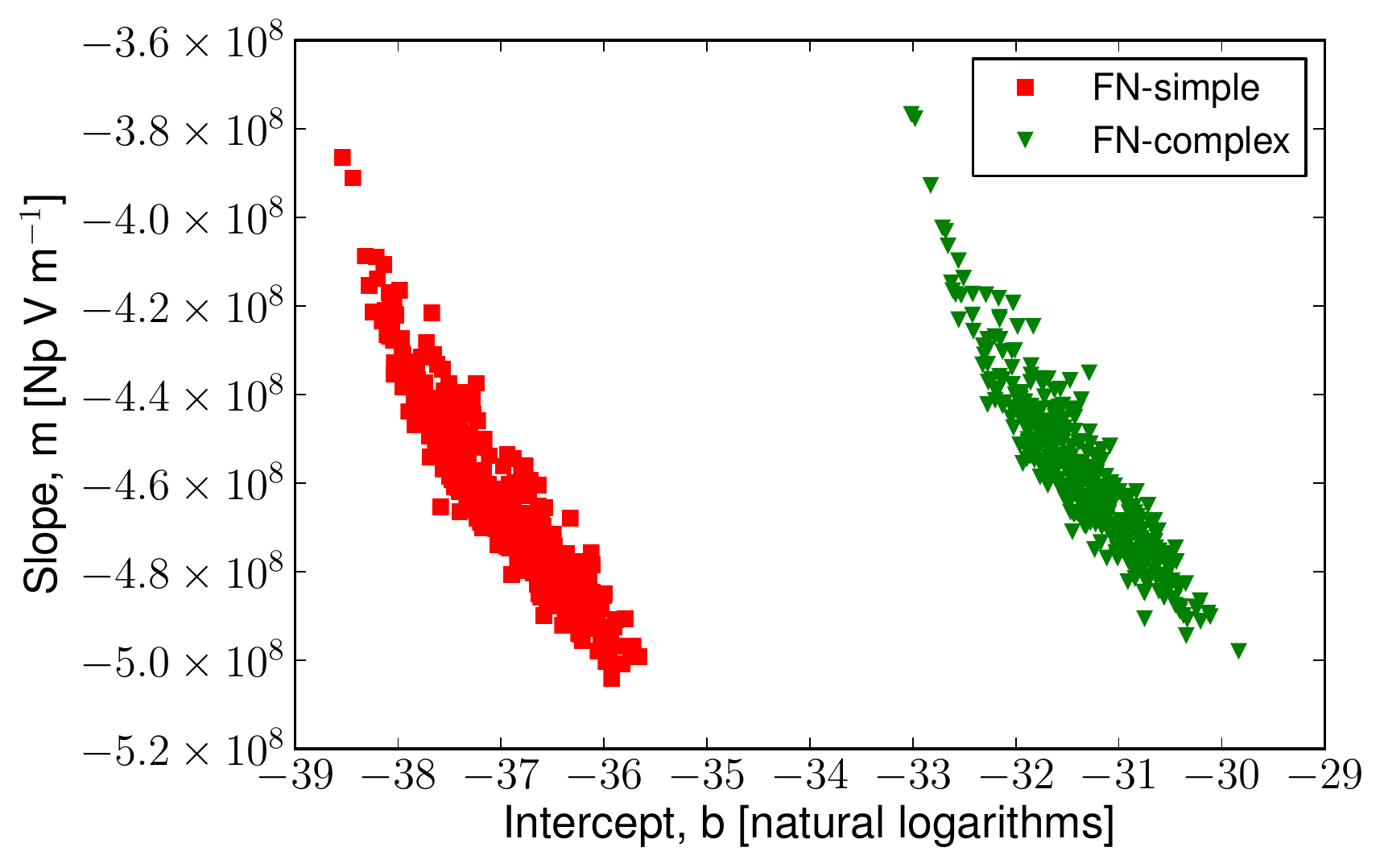}
  \caption{SK-plot for log-normal radial distribution, log-normal
    height, and constant work function. Comparing the use of the simple
    FN-equation vs. the more complex one. Each simulation used 400
    samples with 1024 tips.}
  \label{fig:FN}
\end{figure}

Similarly, the simple FN-equation yields the same results compared to the
more realistic one, as seen in Fig.~\ref{fig:FN}.

\begin{figure}[ht!]
  \centering
  \includegraphics[width=\linewidth]{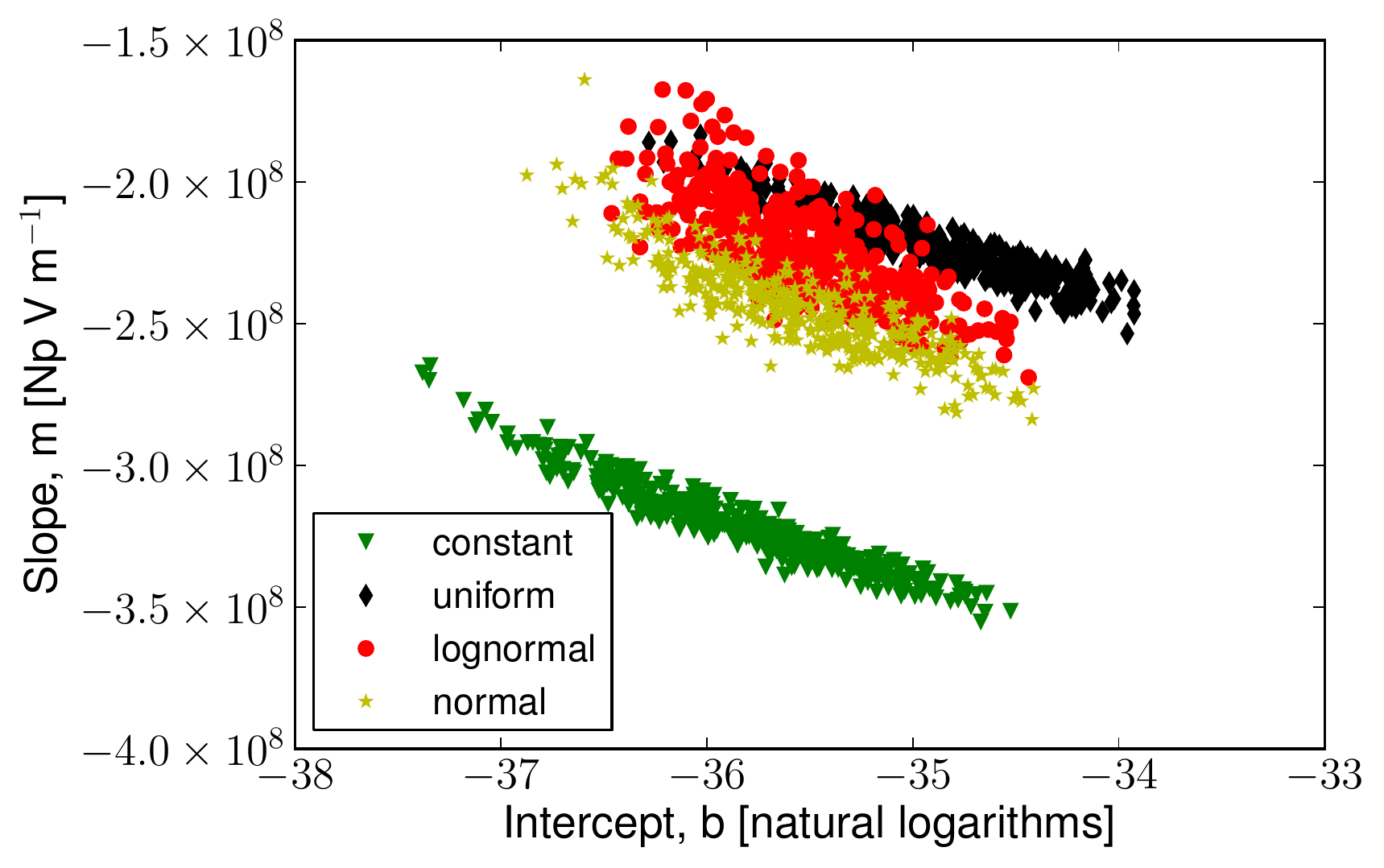}
  \caption{SK-plot for log-normal radial distribution and constant
    work function. Varying the type of distribution for the height.
    Each simulation used 400 samples with 1024 tips.}
  \label{fig:H-dist}
\end{figure}

\begin{figure}[ht!]
  \centering
  \includegraphics[width=\linewidth]{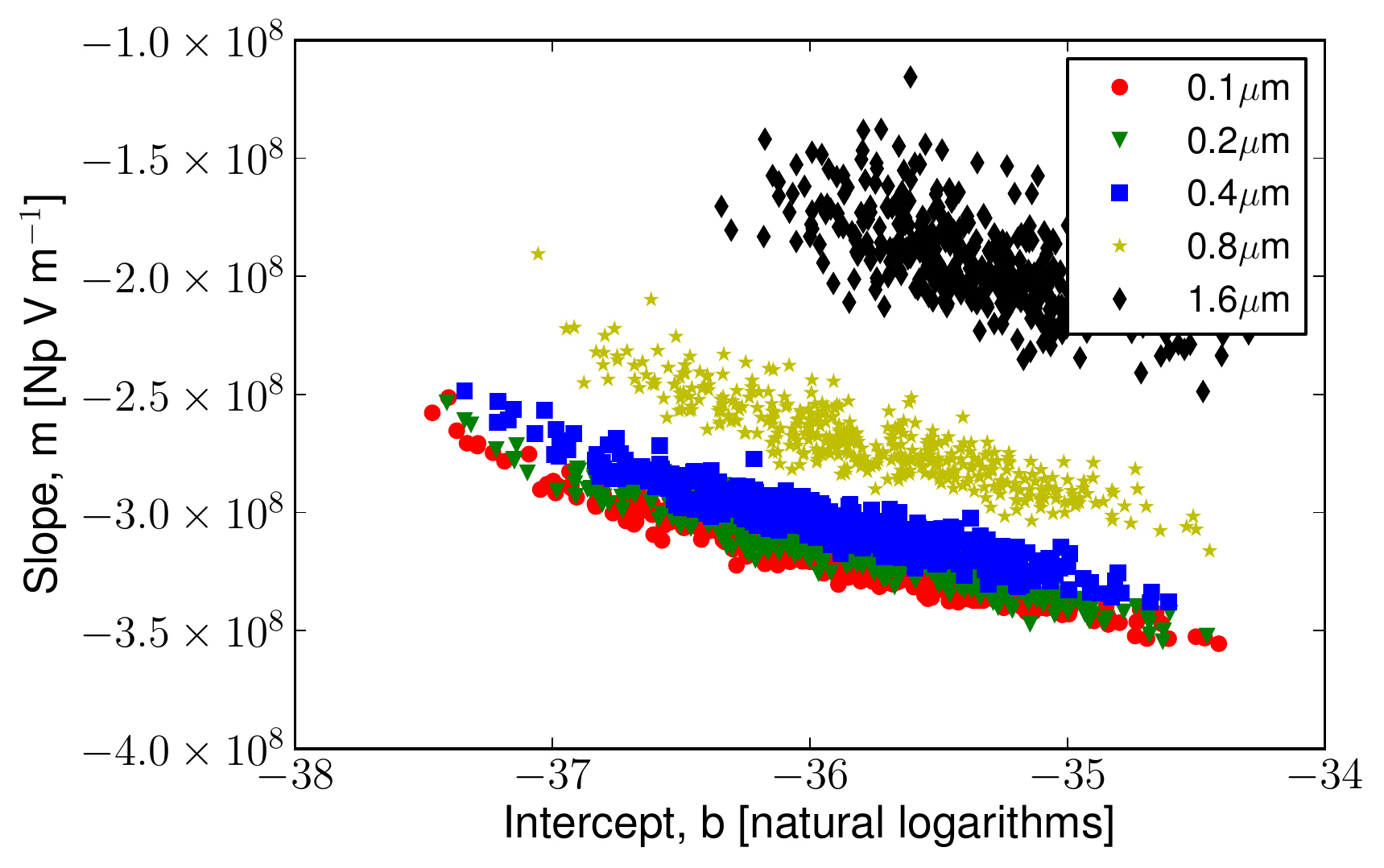}
  \caption{SK-plot for log-normal radial distribution, log-normal
    height, and constant work function. Varying the standard deviation
    of the height distribution. Each simulation used 400 samples with
    1024 tips.}
  \label{fig:H}
\end{figure}

In addition, using different distributions for the height does
not influence the shape of the SK-plots, as can be seen in Fig.~\ref{fig:H-dist}.
In Fig.~\ref{fig:H}, different SK-plots
are shown for a log-normal height distributions with different standard
deviations. The width of the resulting plot and the
position vary, but the distributions along the length of the SK-plot seem to be
dominated by the radial distribution for this parameter space.

\begin{figure}[ht!]
  \centering
  \includegraphics[width=\linewidth]{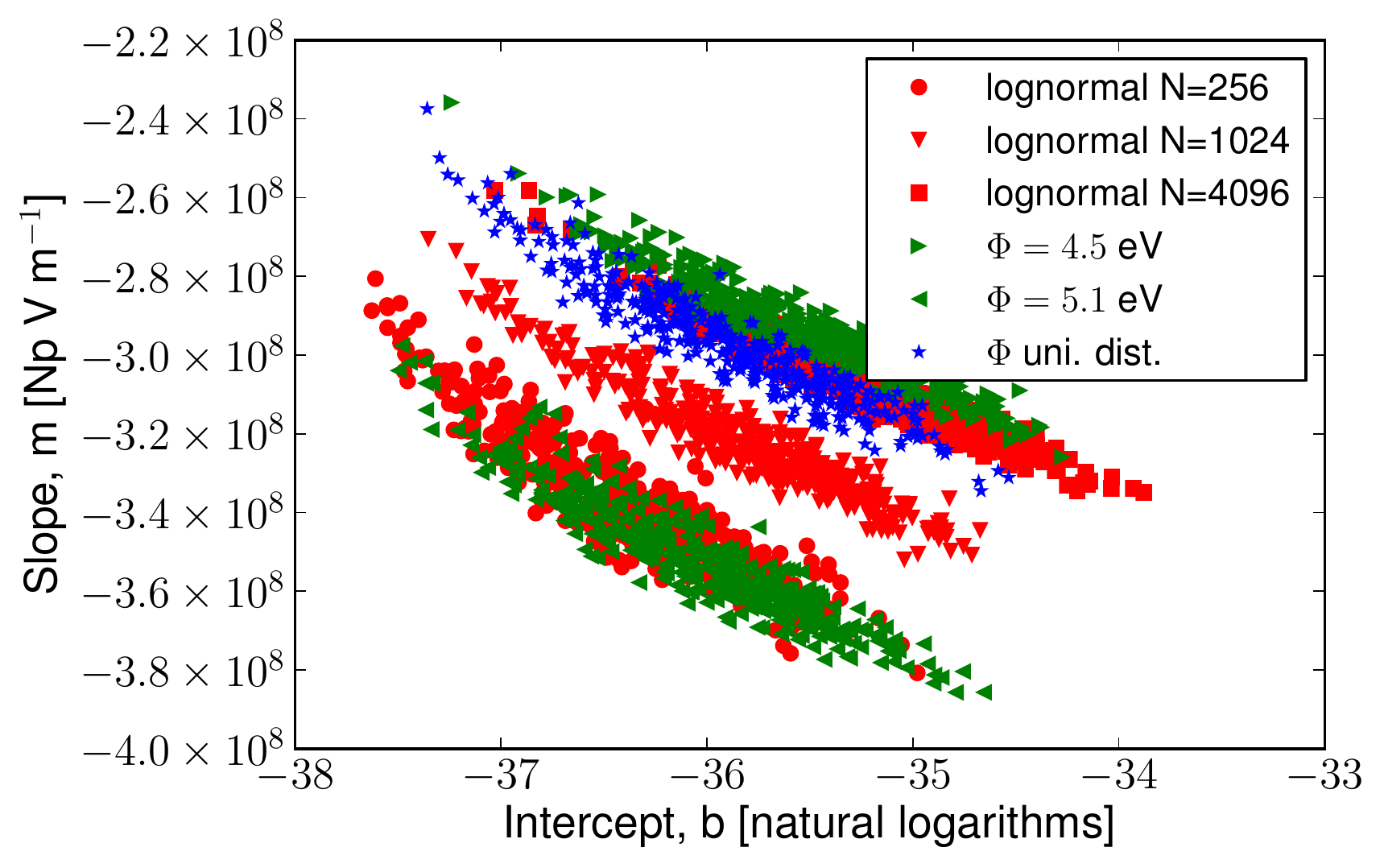}
  \caption{Comparing different sample sizes with a change in work
    function. All simulations use a constant value for the height and
    the work function, unless otherwise noted. The simulation used 400
    samples for each case and 1024 tips, unless otherwise noted.}
  \label{fig:phi}
\end{figure}

Finally, the distribution of different work function values
with SK-plots for different sample sizes are compared. In Fig.~\ref{fig:phi}, one can
see that it is impossible to distinguish a work function change
of \unit[0.3]{eV} and a change in sample size of a factor of $4$.
A distribution of work function values also only slightly influences the
shape of the SK-plot. The resulting SK-plot for a work function that is not
constant across a sample is wider than that for a constant work
function sample, and tends to be similar in values to a higher work function sample.

\section{Conclusions}
As shown, experimental electron field-emission data obtained using tip
arrays can be reproduced assuming a simple model and a log-normal
distribution of tip
radii. The model can explain the spread observed in the SK-plots and
the almost linear relationship observed, as well as
the fact that the data points appear randomly along the
line of the plot. The effect seems to be purely statistical in nature due to the
finite number of tips on each sample and the fact that a log-normal
distribution produces a few very sharp tips leading to a larger
distribution in the emitted
current. Since the calculations were carried out for arrays of tips,
it is not immediately clear how the almost linear form of an SK-plot
can also be reproduced for single tips.\cite{Kawasaki2010} One can speculate
that this is due to statistical variations in nano-protrusions on the tip
surface, which for a small number of protrusions has been shown to
result in a similar effect.\cite{Charbonnier2004} Similarly, for
repeated measurements on a single sample, the observation of nearly linear
SK-plots indicates that the radial distributions change over time while still following a log-normal or similar distribution.

The simulations also show that a change in work function will shift the
SK-plot, supporting analyses conducted elsewhere, as for example in
the paper of Gotoh \textit{et al.}\cite{Gotoh2007b}
However, the simulations also show that changes in the number of emitters per
sample give similar results to those obtained when varying the work function. Therefore, if the number of emitters is
not well known, no conclusion about the work function can be made.

It will not be possible to extract all distribution parameters from
measured data using this model, since changes in several parameters
lead to the same effect, for example, changes in work function
vs. changes in sample size, or changes in standard deviation of the
height distribution vs. a distribution of work function values.  This
is similar to the known fact that one cannot extract the field enhancement
factor and the work function reliably at the same time from a single
FN-plot. However, when assumptions can be made for several of the parameters one should be able to extract information from the
sample by creating SK-plots with good statistics. Here, one has to
make sure that the correction factors are used, since these can
create large offsets in the absolute values of slope and intercept, as
shown. The extraction of
distribution parameters using this approach will be explored in a
future publication.

\section*{Acknowledgements}
The author would like to thank Thomas Schenkel, Christoph Weis, and
Frances Allen for helpful discussions.

This work was supported by the Office of Proliferation Detection (DNN
R\&D) of the US Department of Energy at the Lawrence Berkeley National
Laboratory under contract number DE-AC02-05CHI1231.

%\section*{References}

%\bibliography{slope-paper}

\begin{thebibliography}{34}%
\makeatletter
\providecommand \@ifxundefined [1]{%
 \@ifx{#1\undefined}
}%
\providecommand \@ifnum [1]{%
 \ifnum #1\expandafter \@firstoftwo
 \else \expandafter \@secondoftwo
 \fi
}%
\providecommand \@ifx [1]{%
 \ifx #1\expandafter \@firstoftwo
 \else \expandafter \@secondoftwo
 \fi
}%
\providecommand \natexlab [1]{#1}%
\providecommand \enquote  [1]{``#1''}%
\providecommand \bibnamefont  [1]{#1}%
\providecommand \bibfnamefont [1]{#1}%
\providecommand \citenamefont [1]{#1}%
\providecommand \href@noop [0]{\@secondoftwo}%
\providecommand \href [0]{\begingroup \@sanitize@url \@href}%
\providecommand \@href[1]{\@@startlink{#1}\@@href}%
\providecommand \@@href[1]{\endgroup#1\@@endlink}%
\providecommand \@sanitize@url [0]{\catcode `\\12\catcode `\$12\catcode
  `\&12\catcode `\#12\catcode `\^12\catcode `\_12\catcode `\%12\relax}%
\providecommand \@@startlink[1]{}%
\providecommand \@@endlink[0]{}%
\providecommand \url  [0]{\begingroup\@sanitize@url \@url }%
\providecommand \@url [1]{\endgroup\@href {#1}{\urlprefix }}%
\providecommand \urlprefix  [0]{URL }%
\providecommand \Eprint [0]{\href }%
\providecommand \doibase [0]{http://dx.doi.org/}%
\providecommand \selectlanguage [0]{\@gobble}%
\providecommand \bibinfo  [0]{\@secondoftwo}%
\providecommand \bibfield  [0]{\@secondoftwo}%
\providecommand \translation [1]{[#1]}%
\providecommand \BibitemOpen [0]{}%
\providecommand \bibitemStop [0]{}%
\providecommand \bibitemNoStop [0]{.\EOS\space}%
\providecommand \EOS [0]{\spacefactor3000\relax}%
\providecommand \BibitemShut  [1]{\csname bibitem#1\endcsname}%
\let\auto@bib@innerbib\@empty
%</preamble>
\bibitem [{\citenamefont {Gaertner}(2012)}]{Gaertner2012}%
  \BibitemOpen
  \bibfield  {author} {\bibinfo {author} {\bibfnamefont {G.}~\bibnamefont
  {Gaertner}},\ }\href {\doibase 10.1116/1.4747705} {\bibfield  {journal}
  {\bibinfo  {journal} {J. Vac. Sci. Technol. B}\ }\textbf {\bibinfo {volume}
  {30}},\ \bibinfo {pages} {060801} (\bibinfo {year} {2012})}\BibitemShut
  {NoStop}%
\bibitem [{\citenamefont {Naranjo}, \citenamefont {Gimzewski},\ and\
  \citenamefont {Putterman}(2005)}]{Naranjo2005}%
  \BibitemOpen
  \bibfield  {author} {\bibinfo {author} {\bibfnamefont {B.}~\bibnamefont
  {Naranjo}}, \bibinfo {author} {\bibfnamefont {J.~K.}\ \bibnamefont
  {Gimzewski}}, \ and\ \bibinfo {author} {\bibfnamefont {S.}~\bibnamefont
  {Putterman}},\ }\href {\doibase 10.1038/nature03575} {\bibfield  {journal}
  {\bibinfo  {journal} {Nature}\ }\textbf {\bibinfo {volume} {434}},\ \bibinfo
  {pages} {1115} (\bibinfo {year} {2005})}\BibitemShut {NoStop}%
\bibitem [{\citenamefont {Resnick}\ \emph {et~al.}(2010)\citenamefont
  {Resnick}, \citenamefont {Holland}, \citenamefont {Schwoebel}, \citenamefont
  {Hertz},\ and\ \citenamefont {Chichester}}]{Resnick20101263}%
  \BibitemOpen
  \bibfield  {author} {\bibinfo {author} {\bibfnamefont {P.}~\bibnamefont
  {Resnick}}, \bibinfo {author} {\bibfnamefont {C.}~\bibnamefont {Holland}},
  \bibinfo {author} {\bibfnamefont {P.}~\bibnamefont {Schwoebel}}, \bibinfo
  {author} {\bibfnamefont {K.}~\bibnamefont {Hertz}}, \ and\ \bibinfo {author}
  {\bibfnamefont {D.}~\bibnamefont {Chichester}},\ }\href {\doibase
  10.1016/j.mee.2009.11.036} {\bibfield  {journal} {\bibinfo  {journal}
  {Microelectronic Engineering}\ }\textbf {\bibinfo {volume} {87}},\ \bibinfo
  {pages} {1263} (\bibinfo {year} {2010})}\BibitemShut {NoStop}%
\bibitem [{\citenamefont {Ellsworth}\ \emph {et~al.}(2013)\citenamefont
  {Ellsworth}, \citenamefont {Tang}, \citenamefont {Falabella}, \citenamefont
  {Naranjo},\ and\ \citenamefont {Putterman}}]{Ellsworth2013}%
  \BibitemOpen
  \bibfield  {author} {\bibinfo {author} {\bibfnamefont {J.~L.}\ \bibnamefont
  {Ellsworth}}, \bibinfo {author} {\bibfnamefont {V.}~\bibnamefont {Tang}},
  \bibinfo {author} {\bibfnamefont {S.}~\bibnamefont {Falabella}}, \bibinfo
  {author} {\bibfnamefont {B.}~\bibnamefont {Naranjo}}, \ and\ \bibinfo
  {author} {\bibfnamefont {S.}~\bibnamefont {Putterman}},\ }\href {\doibase
  10.1063/1.4802305} {\bibfield  {journal} {\bibinfo  {journal} {AIP Conf.
  Proc.}\ }\textbf {\bibinfo {volume} {1525}},\ \bibinfo {pages} {128}
  (\bibinfo {year} {2013})}\BibitemShut {NoStop}%
\bibitem [{\citenamefont {Persaud}\ \emph {et~al.}(2012)\citenamefont
  {Persaud}, \citenamefont {Waldmann}, \citenamefont {Kapadia}, \citenamefont
  {Takei}, \citenamefont {Javey},\ and\ \citenamefont
  {Schenkel}}]{Persaud:RoSI-83-02B312}%
  \BibitemOpen
  \bibfield  {author} {\bibinfo {author} {\bibfnamefont {A.}~\bibnamefont
  {Persaud}}, \bibinfo {author} {\bibfnamefont {O.}~\bibnamefont {Waldmann}},
  \bibinfo {author} {\bibfnamefont {R.}~\bibnamefont {Kapadia}}, \bibinfo
  {author} {\bibfnamefont {K.}~\bibnamefont {Takei}}, \bibinfo {author}
  {\bibfnamefont {A.}~\bibnamefont {Javey}}, \ and\ \bibinfo {author}
  {\bibfnamefont {T.}~\bibnamefont {Schenkel}},\ }\href {\doibase
  10.1063/1.3672437} {\bibfield  {journal} {\bibinfo  {journal} {Rev. Sci.
  Instrum.}\ }\textbf {\bibinfo {volume} {83}},\ \bibinfo {pages} {02B312}
  (\bibinfo {year} {2012})}\BibitemShut {NoStop}%
\bibitem [{\citenamefont {Persaud}\ \emph {et~al.}(2011)\citenamefont
  {Persaud}, \citenamefont {Allen}, \citenamefont {Dickinson}, \citenamefont
  {Schenkel}, \citenamefont {Kapadia}, \citenamefont {Takei},\ and\
  \citenamefont {Javey}}]{Persaud:JVSTB-29-02B107}%
  \BibitemOpen
  \bibfield  {author} {\bibinfo {author} {\bibfnamefont {A.}~\bibnamefont
  {Persaud}}, \bibinfo {author} {\bibfnamefont {I.}~\bibnamefont {Allen}},
  \bibinfo {author} {\bibfnamefont {M.~R.}\ \bibnamefont {Dickinson}}, \bibinfo
  {author} {\bibfnamefont {T.}~\bibnamefont {Schenkel}}, \bibinfo {author}
  {\bibfnamefont {R.}~\bibnamefont {Kapadia}}, \bibinfo {author} {\bibfnamefont
  {K.}~\bibnamefont {Takei}}, \ and\ \bibinfo {author} {\bibfnamefont
  {A.}~\bibnamefont {Javey}},\ }\href {\doibase 10.1116/1.3531929} {\bibfield
  {journal} {\bibinfo  {journal} {J. Vac. Sci. Technol. B}\ }\textbf {\bibinfo
  {volume} {29}},\ \bibinfo {pages} {02B107} (\bibinfo {year}
  {2011})}\BibitemShut {NoStop}%
\bibitem [{\citenamefont {Waldmann}\ \emph {et~al.}(2013)\citenamefont
  {Waldmann}, \citenamefont {Persaud}, \citenamefont {Kapadia}, \citenamefont
  {Takei}, \citenamefont {Allen}, \citenamefont {Javey},\ and\ \citenamefont
  {Schenkel}}]{Waldmann:TSF-534-48}%
  \BibitemOpen
  \bibfield  {author} {\bibinfo {author} {\bibfnamefont {O.}~\bibnamefont
  {Waldmann}}, \bibinfo {author} {\bibfnamefont {A.}~\bibnamefont {Persaud}},
  \bibinfo {author} {\bibfnamefont {R.}~\bibnamefont {Kapadia}}, \bibinfo
  {author} {\bibfnamefont {K.}~\bibnamefont {Takei}}, \bibinfo {author}
  {\bibfnamefont {F.~I.}\ \bibnamefont {Allen}}, \bibinfo {author}
  {\bibfnamefont {A.}~\bibnamefont {Javey}}, \ and\ \bibinfo {author}
  {\bibfnamefont {T.}~\bibnamefont {Schenkel}},\ }\href {\doibase
  10.1016/j.tsf.2013.02.053} {\bibfield  {journal} {\bibinfo  {journal} {Thin
  Solid Films}\ }\textbf {\bibinfo {volume} {534}},\ \bibinfo {pages} {488}
  (\bibinfo {year} {2013})}\BibitemShut {NoStop}%
\bibitem [{\citenamefont {Fowler}\ and\ \citenamefont
  {Nordheim}(1928)}]{Fowler1928}%
  \BibitemOpen
  \bibfield  {author} {\bibinfo {author} {\bibfnamefont {R.~H.}\ \bibnamefont
  {Fowler}}\ and\ \bibinfo {author} {\bibfnamefont {L.}~\bibnamefont
  {Nordheim}},\ }\href {\doibase 10.1098/rspa.1928.0091} {\bibfield  {journal}
  {\bibinfo  {journal} {Proc. R. Soc. London. Ser. A}\ }\textbf {\bibinfo
  {volume} {119}},\ \bibinfo {pages} {173} (\bibinfo {year}
  {1928})}\BibitemShut {NoStop}%
\bibitem [{\citenamefont {Forbes}, \citenamefont {Fischer},\ and\ \citenamefont
  {Mousa}(2013)}]{Forbes2013}%
  \BibitemOpen
  \bibfield  {author} {\bibinfo {author} {\bibfnamefont {R.~G.}\ \bibnamefont
  {Forbes}}, \bibinfo {author} {\bibfnamefont {A.}~\bibnamefont {Fischer}}, \
  and\ \bibinfo {author} {\bibfnamefont {M.~S.}\ \bibnamefont {Mousa}},\ }\href
  {\doibase 10.1116/1.4765080} {\bibfield  {journal} {\bibinfo  {journal} {J.
  Vac. Sci. Technol. B}\ }\textbf {\bibinfo {volume} {31}},\ \bibinfo {pages}
  {02B103} (\bibinfo {year} {2013})}\BibitemShut {NoStop}%
\bibitem [{\citenamefont {Mackie}\ \emph {et~al.}(2003)\citenamefont {Mackie},
  \citenamefont {Southall}, \citenamefont {Xie}, \citenamefont {Cabe},
  \citenamefont {Charbonnier},\ and\ \citenamefont {McClelland}}]{Mackie2003}%
  \BibitemOpen
  \bibfield  {author} {\bibinfo {author} {\bibfnamefont {W.~A.}\ \bibnamefont
  {Mackie}}, \bibinfo {author} {\bibfnamefont {L.~A.}\ \bibnamefont
  {Southall}}, \bibinfo {author} {\bibfnamefont {T.}~\bibnamefont {Xie}},
  \bibinfo {author} {\bibfnamefont {G.~L.}\ \bibnamefont {Cabe}}, \bibinfo
  {author} {\bibfnamefont {F.~M.}\ \bibnamefont {Charbonnier}}, \ and\ \bibinfo
  {author} {\bibfnamefont {P.~H.}\ \bibnamefont {McClelland}},\ }\href
  {\doibase 10.1116/1.1576764} {\bibfield  {journal} {\bibinfo  {journal} {J.
  Vac. Sci. Technol. B}\ }\textbf {\bibinfo {volume} {21}},\ \bibinfo {pages}
  {1574} (\bibinfo {year} {2003})}\BibitemShut {NoStop}%
\bibitem [{\citenamefont {Charbonnier}, \citenamefont {Southall},\ and\
  \citenamefont {Mackie}(2004)}]{Charbonnier2004}%
  \BibitemOpen
  \bibfield  {author} {\bibinfo {author} {\bibfnamefont {F.~M.}\ \bibnamefont
  {Charbonnier}}, \bibinfo {author} {\bibfnamefont {L.~A.}\ \bibnamefont
  {Southall}}, \ and\ \bibinfo {author} {\bibfnamefont {W.~A.}\ \bibnamefont
  {Mackie}},\ }\href {\doibase 10.1116/1.1760753} {\bibfield  {journal}
  {\bibinfo  {journal} {J. Vac. Sci. Technol. B}\ }\textbf {\bibinfo {volume}
  {22}},\ \bibinfo {pages} {1643} (\bibinfo {year} {2004})}\BibitemShut
  {NoStop}%
\bibitem [{\citenamefont {Gotoh}\ \emph
  {et~al.}(2007{\natexlab{a}})\citenamefont {Gotoh}, \citenamefont {Kawamura},
  \citenamefont {Niiya}, \citenamefont {Ishibashi}, \citenamefont {Nicolaescu},
  \citenamefont {Tsuji}, \citenamefont {Ishikawa}, \citenamefont {Hosono},
  \citenamefont {Nakata},\ and\ \citenamefont {Okuda}}]{Gotoh2007a}%
  \BibitemOpen
  \bibfield  {author} {\bibinfo {author} {\bibfnamefont {Y.}~\bibnamefont
  {Gotoh}}, \bibinfo {author} {\bibfnamefont {Y.}~\bibnamefont {Kawamura}},
  \bibinfo {author} {\bibfnamefont {T.}~\bibnamefont {Niiya}}, \bibinfo
  {author} {\bibfnamefont {T.}~\bibnamefont {Ishibashi}}, \bibinfo {author}
  {\bibfnamefont {D.}~\bibnamefont {Nicolaescu}}, \bibinfo {author}
  {\bibfnamefont {H.}~\bibnamefont {Tsuji}}, \bibinfo {author} {\bibfnamefont
  {J.}~\bibnamefont {Ishikawa}}, \bibinfo {author} {\bibfnamefont
  {A.}~\bibnamefont {Hosono}}, \bibinfo {author} {\bibfnamefont
  {S.}~\bibnamefont {Nakata}}, \ and\ \bibinfo {author} {\bibfnamefont
  {S.}~\bibnamefont {Okuda}},\ }\href {\doibase 10.1063/1.2740199} {\bibfield
  {journal} {\bibinfo  {journal} {Appl. Phys. Lett.}\ }\textbf {\bibinfo
  {volume} {90}},\ \bibinfo {pages} {203107} (\bibinfo {year}
  {2007}{\natexlab{a}})}\BibitemShut {NoStop}%
\bibitem [{\citenamefont {Kawasaki}\ \emph {et~al.}(2010)\citenamefont
  {Kawasaki}, \citenamefont {He}, \citenamefont {Gotoh}, \citenamefont
  {Tsuji},\ and\ \citenamefont {Ishikawa}}]{Kawasaki2010}%
  \BibitemOpen
  \bibfield  {author} {\bibinfo {author} {\bibfnamefont {M.}~\bibnamefont
  {Kawasaki}}, \bibinfo {author} {\bibfnamefont {Z.}~\bibnamefont {He}},
  \bibinfo {author} {\bibfnamefont {Y.}~\bibnamefont {Gotoh}}, \bibinfo
  {author} {\bibfnamefont {H.}~\bibnamefont {Tsuji}}, \ and\ \bibinfo {author}
  {\bibfnamefont {J.}~\bibnamefont {Ishikawa}},\ }\href {\doibase
  10.1116/1.3325835} {\bibfield  {journal} {\bibinfo  {journal} {J. Vac. Sci.
  Technol. B}\ }\textbf {\bibinfo {volume} {28}},\ \bibinfo {pages} {C2A77}
  (\bibinfo {year} {2010})}\BibitemShut {NoStop}%
\bibitem [{\citenamefont {Gomer}(1994)}]{Gomer1994}%
  \BibitemOpen
  \bibfield  {author} {\bibinfo {author} {\bibfnamefont {R.}~\bibnamefont
  {Gomer}},\ }\href {\doibase 10.1016/0039-6028(94)90651-3} {\bibfield
  {journal} {\bibinfo  {journal} {Surf. Sci.}\ }\textbf {\bibinfo {volume}
  {299-300}},\ \bibinfo {pages} {129} (\bibinfo {year} {1994})}\BibitemShut
  {NoStop}%
\bibitem [{\citenamefont {Forbes}(2012)}]{Forbes2012}%
  \BibitemOpen
  \bibfield  {author} {\bibinfo {author} {\bibfnamefont {R.~G.}\ \bibnamefont
  {Forbes}},\ }\href {\doibase 10.1088/0957-4484/23/9/095706} {\bibfield
  {journal} {\bibinfo  {journal} {Nanotechnology}\ }\textbf {\bibinfo {volume}
  {23}},\ \bibinfo {pages} {095706} (\bibinfo {year} {2012})}\BibitemShut
  {NoStop}%
\bibitem [{\citenamefont {Edgcombe}\ and\ \citenamefont
  {Valdre}(2001)}]{Edgcombe2001}%
  \BibitemOpen
  \bibfield  {author} {\bibinfo {author} {\bibfnamefont {C.~J.}\ \bibnamefont
  {Edgcombe}}\ and\ \bibinfo {author} {\bibfnamefont {U.}~\bibnamefont
  {Valdre}},\ }\href {\doibase 10.1046/j.1365-2818.2001.00890.x} {\bibfield
  {journal} {\bibinfo  {journal} {Journal of Microscopy}\ }\textbf {\bibinfo
  {volume} {203}},\ \bibinfo {pages} {188} (\bibinfo {year}
  {2001})}\BibitemShut {NoStop}%
\bibitem [{\citenamefont {Bonard}\ \emph {et~al.}(2001)\citenamefont {Bonard},
  \citenamefont {Weiss}, \citenamefont {Kind}, \citenamefont {St\"{o}ckli},
  \citenamefont {Forr\'{o}}, \citenamefont {Kern},\ and\ \citenamefont
  {Ch\^atelain}}]{Bonard2001}%
  \BibitemOpen
  \bibfield  {author} {\bibinfo {author} {\bibfnamefont {J.}~\bibnamefont
  {Bonard}}, \bibinfo {author} {\bibfnamefont {N.}~\bibnamefont {Weiss}},
  \bibinfo {author} {\bibfnamefont {H.}~\bibnamefont {Kind}}, \bibinfo {author}
  {\bibfnamefont {T.}~\bibnamefont {St\"{o}ckli}}, \bibinfo {author}
  {\bibfnamefont {L.}~\bibnamefont {Forr\'{o}}}, \bibinfo {author}
  {\bibfnamefont {K.}~\bibnamefont {Kern}}, \ and\ \bibinfo {author}
  {\bibfnamefont {A.}~\bibnamefont {Ch\^atelain}},\ }\href@noop {} {\bibfield
  {journal} {\bibinfo  {journal} {Advanced materials}\ }\textbf {\bibinfo
  {volume} {13}},\ \bibinfo {pages} {184} (\bibinfo {year} {2001})}\BibitemShut
  {NoStop}%
\bibitem [{\citenamefont {Jo}\ \emph {et~al.}(2003)\citenamefont {Jo},
  \citenamefont {Tu}, \citenamefont {Huang}, \citenamefont {Carnahan},
  \citenamefont {Wang},\ and\ \citenamefont {Ren}}]{Jo2003}%
  \BibitemOpen
  \bibfield  {author} {\bibinfo {author} {\bibfnamefont {S.~H.}\ \bibnamefont
  {Jo}}, \bibinfo {author} {\bibfnamefont {Y.}~\bibnamefont {Tu}}, \bibinfo
  {author} {\bibfnamefont {Z.~P.}\ \bibnamefont {Huang}}, \bibinfo {author}
  {\bibfnamefont {D.~L.}\ \bibnamefont {Carnahan}}, \bibinfo {author}
  {\bibfnamefont {D.~Z.}\ \bibnamefont {Wang}}, \ and\ \bibinfo {author}
  {\bibfnamefont {Z.~F.}\ \bibnamefont {Ren}},\ }\href {\doibase
  10.1063/1.1576310} {\bibfield  {journal} {\bibinfo  {journal} {Appl. Phys.
  Lett.}\ }\textbf {\bibinfo {volume} {82}},\ \bibinfo {pages} {3520} (\bibinfo
  {year} {2003})}\BibitemShut {NoStop}%
\bibitem [{\citenamefont {Ishikawa}\ \emph {et~al.}(1993)\citenamefont
  {Ishikawa}, \citenamefont {Tsuji}, \citenamefont {Gotoh}, \citenamefont
  {Sasaki}, \citenamefont {Kaneko}, \citenamefont {Nagao},\ and\ \citenamefont
  {Inoue}}]{Ishikawa1993}%
  \BibitemOpen
  \bibfield  {author} {\bibinfo {author} {\bibfnamefont {J.}~\bibnamefont
  {Ishikawa}}, \bibinfo {author} {\bibfnamefont {H.}~\bibnamefont {Tsuji}},
  \bibinfo {author} {\bibfnamefont {Y.}~\bibnamefont {Gotoh}}, \bibinfo
  {author} {\bibfnamefont {T.}~\bibnamefont {Sasaki}}, \bibinfo {author}
  {\bibfnamefont {T.}~\bibnamefont {Kaneko}}, \bibinfo {author} {\bibfnamefont
  {M.}~\bibnamefont {Nagao}}, \ and\ \bibinfo {author} {\bibfnamefont
  {K.}~\bibnamefont {Inoue}},\ }\href {\doibase 10.1116/1.586870} {\bibfield
  {journal} {\bibinfo  {journal} {J. Vac. Sci. Technol. B}\ }\textbf {\bibinfo
  {volume} {11}},\ \bibinfo {pages} {403} (\bibinfo {year} {1993})}\BibitemShut
  {NoStop}%
\bibitem [{\citenamefont {Gotoh}\ \emph {et~al.}(1996)\citenamefont {Gotoh},
  \citenamefont {Nagao}, \citenamefont {Matsubara}, \citenamefont {Inoue},
  \citenamefont {Tsuji},\ and\ \citenamefont {Ishikawa}}]{Gotoh1996}%
  \BibitemOpen
  \bibfield  {author} {\bibinfo {author} {\bibfnamefont {Y.}~\bibnamefont
  {Gotoh}}, \bibinfo {author} {\bibfnamefont {M.}~\bibnamefont {Nagao}},
  \bibinfo {author} {\bibfnamefont {M.}~\bibnamefont {Matsubara}}, \bibinfo
  {author} {\bibfnamefont {K.}~\bibnamefont {Inoue}}, \bibinfo {author}
  {\bibfnamefont {H.}~\bibnamefont {Tsuji}}, \ and\ \bibinfo {author}
  {\bibfnamefont {J.}~\bibnamefont {Ishikawa}},\ }\href {\doibase
  10.1143/JJAP.35.L1297} {\bibfield  {journal} {\bibinfo  {journal} {Jpn. J.
  Appl. Phys.}\ }\textbf {\bibinfo {volume} {35}},\ \bibinfo {pages} {L1297}
  (\bibinfo {year} {1996})}\BibitemShut {NoStop}%
\bibitem [{\citenamefont {da~Rocha}\ \emph {et~al.}(2008)\citenamefont
  {da~Rocha}, \citenamefont {Santos}, \citenamefont {de~Paulo}, \citenamefont
  {Hering}, \citenamefont {Engelsen}, \citenamefont {Vuolo}, \citenamefont
  {Mammana},\ and\ \citenamefont {Mammana}}]{DaRocha2008}%
  \BibitemOpen
  \bibfield  {author} {\bibinfo {author} {\bibfnamefont {M.}~\bibnamefont
  {da~Rocha}}, \bibinfo {author} {\bibfnamefont {T.}~\bibnamefont {Santos}},
  \bibinfo {author} {\bibfnamefont {A.}~\bibnamefont {de~Paulo}}, \bibinfo
  {author} {\bibfnamefont {V.}~\bibnamefont {Hering}}, \bibinfo {author}
  {\bibfnamefont {D.~D.}\ \bibnamefont {Engelsen}}, \bibinfo {author}
  {\bibfnamefont {J.}~\bibnamefont {Vuolo}}, \bibinfo {author} {\bibfnamefont
  {S.}~\bibnamefont {Mammana}}, \ and\ \bibinfo {author} {\bibfnamefont
  {V.}~\bibnamefont {Mammana}},\ }\href {\doibase 10.1016/j.apsusc.2007.07.172}
  {\bibfield  {journal} {\bibinfo  {journal} {Applied Surface Science}\
  }\textbf {\bibinfo {volume} {254}},\ \bibinfo {pages} {1859} (\bibinfo {year}
  {2008})}\BibitemShut {NoStop}%
\bibitem [{\citenamefont {Nicolaescu}\ \emph {et~al.}(2001)\citenamefont
  {Nicolaescu}, \citenamefont {Filip}, \citenamefont {Itoh},\ and\
  \citenamefont {Okuyama}}]{Nicolaescu2001}%
  \BibitemOpen
  \bibfield  {author} {\bibinfo {author} {\bibfnamefont {D.}~\bibnamefont
  {Nicolaescu}}, \bibinfo {author} {\bibfnamefont {V.}~\bibnamefont {Filip}},
  \bibinfo {author} {\bibfnamefont {J.}~\bibnamefont {Itoh}}, \ and\ \bibinfo
  {author} {\bibfnamefont {F.}~\bibnamefont {Okuyama}},\ }\href {\doibase
  10.1143/JJAP.40.4802} {\bibfield  {journal} {\bibinfo  {journal} {Jpn. J.
  Appl. Phys.}\ }\textbf {\bibinfo {volume} {40}},\ \bibinfo {pages} {4802}
  (\bibinfo {year} {2001})}\BibitemShut {NoStop}%
\bibitem [{\citenamefont {Gotoh}\ \emph
  {et~al.}(2007{\natexlab{b}})\citenamefont {Gotoh}, \citenamefont {Mukai},
  \citenamefont {Kawamura}, \citenamefont {Tsuji},\ and\ \citenamefont
  {Ishikawa}}]{Gotoh2007b}%
  \BibitemOpen
  \bibfield  {author} {\bibinfo {author} {\bibfnamefont {Y.}~\bibnamefont
  {Gotoh}}, \bibinfo {author} {\bibfnamefont {K.}~\bibnamefont {Mukai}},
  \bibinfo {author} {\bibfnamefont {Y.}~\bibnamefont {Kawamura}}, \bibinfo
  {author} {\bibfnamefont {H.}~\bibnamefont {Tsuji}}, \ and\ \bibinfo {author}
  {\bibfnamefont {J.}~\bibnamefont {Ishikawa}},\ }\href {\doibase
  10.1116/1.2433950} {\bibfield  {journal} {\bibinfo  {journal} {J. Vac. Sci.
  Technol. B}\ }\textbf {\bibinfo {volume} {25}},\ \bibinfo {pages} {508}
  (\bibinfo {year} {2007}{\natexlab{b}})}\BibitemShut {NoStop}%
\bibitem [{\citenamefont {Gotoh}, \citenamefont {Tsuji},\ and\ \citenamefont
  {Ishikawa}(2001)}]{Gotoh2001}%
  \BibitemOpen
  \bibfield  {author} {\bibinfo {author} {\bibfnamefont {Y.}~\bibnamefont
  {Gotoh}}, \bibinfo {author} {\bibfnamefont {H.}~\bibnamefont {Tsuji}}, \ and\
  \bibinfo {author} {\bibfnamefont {J.}~\bibnamefont {Ishikawa}},\ }\href
  {\doibase 10.1016/S0304-3991(01)00117-6} {\bibfield  {journal} {\bibinfo
  {journal} {Ultramicroscopy}\ }\textbf {\bibinfo {volume} {89}},\ \bibinfo
  {pages} {63} (\bibinfo {year} {2001})}\BibitemShut {NoStop}%
\bibitem [{\citenamefont {Gotoh}\ \emph {et~al.}(2004)\citenamefont {Gotoh},
  \citenamefont {Nagao}, \citenamefont {Nozaki}, \citenamefont {Utsumi},
  \citenamefont {Inoue}, \citenamefont {Nakatani}, \citenamefont {Sakashita},
  \citenamefont {Betsui}, \citenamefont {Tsuji},\ and\ \citenamefont
  {Ishikawa}}]{Gotoh2004}%
  \BibitemOpen
  \bibfield  {author} {\bibinfo {author} {\bibfnamefont {Y.}~\bibnamefont
  {Gotoh}}, \bibinfo {author} {\bibfnamefont {M.}~\bibnamefont {Nagao}},
  \bibinfo {author} {\bibfnamefont {D.}~\bibnamefont {Nozaki}}, \bibinfo
  {author} {\bibfnamefont {K.}~\bibnamefont {Utsumi}}, \bibinfo {author}
  {\bibfnamefont {K.}~\bibnamefont {Inoue}}, \bibinfo {author} {\bibfnamefont
  {T.}~\bibnamefont {Nakatani}}, \bibinfo {author} {\bibfnamefont
  {T.}~\bibnamefont {Sakashita}}, \bibinfo {author} {\bibfnamefont
  {K.}~\bibnamefont {Betsui}}, \bibinfo {author} {\bibfnamefont
  {H.}~\bibnamefont {Tsuji}}, \ and\ \bibinfo {author} {\bibfnamefont
  {J.}~\bibnamefont {Ishikawa}},\ }\href {\doibase 10.1063/1.1635655}
  {\bibfield  {journal} {\bibinfo  {journal} {J. Appl. Phys.}\ }\textbf
  {\bibinfo {volume} {95}},\ \bibinfo {pages} {1537} (\bibinfo {year}
  {2004})}\BibitemShut {NoStop}%
\bibitem [{\citenamefont {Nicolaescu}\ \emph {et~al.}(2003)\citenamefont
  {Nicolaescu}, \citenamefont {Nagao}, \citenamefont {Filip}, \citenamefont
  {Kanemaru},\ and\ \citenamefont {Itoh}}]{Nicolaescu2003}%
  \BibitemOpen
  \bibfield  {author} {\bibinfo {author} {\bibfnamefont {D.}~\bibnamefont
  {Nicolaescu}}, \bibinfo {author} {\bibfnamefont {M.}~\bibnamefont {Nagao}},
  \bibinfo {author} {\bibfnamefont {V.}~\bibnamefont {Filip}}, \bibinfo
  {author} {\bibfnamefont {S.}~\bibnamefont {Kanemaru}}, \ and\ \bibinfo
  {author} {\bibfnamefont {J.}~\bibnamefont {Itoh}},\ }\href {\doibase
  10.1116/1.1593641} {\bibfield  {journal} {\bibinfo  {journal} {J. Vac. Sci.
  Technol. B}\ }\textbf {\bibinfo {volume} {21}},\ \bibinfo {pages} {1550}
  (\bibinfo {year} {2003})}\BibitemShut {NoStop}%
\bibitem [{\citenamefont {Nicolaescu}\ \emph {et~al.}(2004)\citenamefont
  {Nicolaescu}, \citenamefont {Sato}, \citenamefont {Nagao}, \citenamefont
  {Filip}, \citenamefont {Kanemaru},\ and\ \citenamefont
  {Itoh}}]{Nicolaescu2004}%
  \BibitemOpen
  \bibfield  {author} {\bibinfo {author} {\bibfnamefont {D.}~\bibnamefont
  {Nicolaescu}}, \bibinfo {author} {\bibfnamefont {T.}~\bibnamefont {Sato}},
  \bibinfo {author} {\bibfnamefont {M.}~\bibnamefont {Nagao}}, \bibinfo
  {author} {\bibfnamefont {V.}~\bibnamefont {Filip}}, \bibinfo {author}
  {\bibfnamefont {S.}~\bibnamefont {Kanemaru}}, \ and\ \bibinfo {author}
  {\bibfnamefont {J.}~\bibnamefont {Itoh}},\ }\href {\doibase
  10.1116/1.1689311} {\bibfield  {journal} {\bibinfo  {journal} {J. Vac. Sci.
  Technol. B}\ }\textbf {\bibinfo {volume} {22}},\ \bibinfo {pages} {1227}
  (\bibinfo {year} {2004})}\BibitemShut {NoStop}%
\bibitem [{\citenamefont {Ding}, \citenamefont {Sha},\ and\ \citenamefont
  {Akinwande}(2002)}]{Ding2002}%
  \BibitemOpen
  \bibfield  {author} {\bibinfo {author} {\bibfnamefont {M.}~\bibnamefont
  {Ding}}, \bibinfo {author} {\bibfnamefont {G.}~\bibnamefont {Sha}}, \ and\
  \bibinfo {author} {\bibfnamefont {A.}~\bibnamefont {Akinwande}},\ }\href
  {\doibase 10.1109/TED.2002.805230} {\bibfield  {journal} {\bibinfo  {journal}
  {IEEE Transactions on Electron Devices}\ }\textbf {\bibinfo {volume} {49}},\
  \bibinfo {pages} {2333} (\bibinfo {year} {2002})}\BibitemShut {NoStop}%
\bibitem [{\citenamefont {Nicolaescu}\ \emph {et~al.}(2006)\citenamefont
  {Nicolaescu}, \citenamefont {Nagao}, \citenamefont {Filip}, \citenamefont
  {Tanoue}, \citenamefont {Kanemaru},\ and\ \citenamefont
  {Itoh}}]{Nicolaescu2006}%
  \BibitemOpen
  \bibfield  {author} {\bibinfo {author} {\bibfnamefont {D.}~\bibnamefont
  {Nicolaescu}}, \bibinfo {author} {\bibfnamefont {M.}~\bibnamefont {Nagao}},
  \bibinfo {author} {\bibfnamefont {V.}~\bibnamefont {Filip}}, \bibinfo
  {author} {\bibfnamefont {H.}~\bibnamefont {Tanoue}}, \bibinfo {author}
  {\bibfnamefont {S.}~\bibnamefont {Kanemaru}}, \ and\ \bibinfo {author}
  {\bibfnamefont {J.}~\bibnamefont {Itoh}},\ }\href {\doibase
  10.1116/1.2184330} {\bibfield  {journal} {\bibinfo  {journal} {J. Vac. Sci.
  Technol. B}\ }\textbf {\bibinfo {volume} {24}},\ \bibinfo {pages} {1045}
  (\bibinfo {year} {2006})}\BibitemShut {NoStop}%
\bibitem [{\citenamefont {Park}, \citenamefont {Lee},\ and\ \citenamefont
  {Koh}(2006)}]{Park2006}%
  \BibitemOpen
  \bibfield  {author} {\bibinfo {author} {\bibfnamefont {K.~H.}\ \bibnamefont
  {Park}}, \bibinfo {author} {\bibfnamefont {S.}~\bibnamefont {Lee}}, \ and\
  \bibinfo {author} {\bibfnamefont {K.~H.}\ \bibnamefont {Koh}},\ }\href
  {\doibase 10.1116/1.2174022} {\bibfield  {journal} {\bibinfo  {journal} {J.
  Vac. Sci. Technol. B}\ }\textbf {\bibinfo {volume} {24}},\ \bibinfo {pages}
  {898} (\bibinfo {year} {2006})}\BibitemShut {NoStop}%
\bibitem [{\citenamefont {Foundation}(95  )}]{python}%
  \BibitemOpen
  \bibfield  {author} {\bibinfo {author} {\bibfnamefont {P.~S.}\ \bibnamefont
  {Foundation}},\ }\href {http://www.python.org} {\enquote {\bibinfo {title}
  {Python language reference, version 2.7},}\ } (\bibinfo {year}
  {1995--})\BibitemShut {NoStop}%
\bibitem [{\citenamefont {Jones}\ \emph {et~al.}(01  )\citenamefont {Jones},
  \citenamefont {Oliphant}, \citenamefont {Peterson} \emph
  {et~al.}}]{Jones2001}%
  \BibitemOpen
  \bibfield  {author} {\bibinfo {author} {\bibfnamefont {E.}~\bibnamefont
  {Jones}}, \bibinfo {author} {\bibfnamefont {T.}~\bibnamefont {Oliphant}},
  \bibinfo {author} {\bibfnamefont {P.}~\bibnamefont {Peterson}},  \emph
  {et~al.},\ }\href {http://www.scipy.org/} {\enquote {\bibinfo {title}
  {{SciPy}: Open source scientific tools for {Python}},}\ } (\bibinfo {year}
  {2001--})\BibitemShut {NoStop}%
\bibitem [{\citenamefont {Hunter}(2007)}]{Hunter:2007}%
  \BibitemOpen
  \bibfield  {author} {\bibinfo {author} {\bibfnamefont {J.~D.}\ \bibnamefont
  {Hunter}},\ }\href@noop {} {\bibfield  {journal} {\bibinfo  {journal}
  {Comput. Sci. Eng.}\ }\textbf {\bibinfo {volume} {9}},\ \bibinfo {pages} {90}
  (\bibinfo {year} {2007})}\BibitemShut {NoStop}%
\bibitem [{\citenamefont {P\'erez}\ and\ \citenamefont
  {Granger}(2007)}]{PER-GRA:2007}%
  \BibitemOpen
  \bibfield  {author} {\bibinfo {author} {\bibfnamefont {F.}~\bibnamefont
  {P\'erez}}\ and\ \bibinfo {author} {\bibfnamefont {B.~E.}\ \bibnamefont
  {Granger}},\ }\href@noop {} {\bibfield  {journal} {\bibinfo  {journal}
  {{C}omput. {S}ci. {E}ng.}\ }\textbf {\bibinfo {volume} {9}},\ \bibinfo
  {pages} {21} (\bibinfo {year} {2007})}\BibitemShut {NoStop}%
\end{thebibliography}

%merlin.mbs aipnum4-1.bst 2010-07-25 4.21a (PWD, AO, DPC) hacked
%Control: key (0)
%Control: author (8) initials jnrlst
%Control: editor formatted (1) identically to author
%Control: production of article title (-1) disabled
%Control: page (0) single
%Control: year (1) truncated
%Control: production of eprint (0) enabled
%

\end{document}